\documentclass[aps,prl,showpacs,amsmath,amssymb,sort&compress,comma,nofootinbib,floatfix,twocolumn,superscriptaddress]{revtex4-1} 
\usepackage{graphicx}
\usepackage{subfigure}  % use for side-by-side figures
\begin{document}

\newcommand{\etal}{{\it et al. }\/}
\newcommand{\gtwid}{\mathrel{\raise.3ex\hbox{$>$\kern-.75em\lower1ex\hbox{$\sim$}}}}
\newcommand{\ltwid}{\mathrel{\raise.3ex\hbox{$<$\kern-.75em\lower1ex\hbox{$\sim$}}}}

\title{Pairing in the Presence of a Pseudogap}

\author{T.A. Maier}
\affiliation{Computer Science and Mathematics Division and Center for Nanophase Materials Sciences,
Oak Ridge National Laboratory, Oak Ridge, Tennessee 37831-6494, USA}

\author{P. Staar}
\affiliation{IBM Research -- Z\"urich, CH-8803 R\"uschlikon, Switzerland}

\author{D.J.~Scalapino}
\affiliation{Department of Physics, University of California, Santa Barbara, CA 93106-9530, USA}

%\date{\today}

%\pacs{71.10.Fd, 03.75.Ss, 74.25.Ha }

\maketitle
%\tableofcontents

{\bf Evidence that the pseudogap (PG) in a near-optimally doped
Bi$_2$Sr$_2$CaCu$_2$O$_{8+\delta}$ sample destroys the BCS logarithmic pairing
instability \cite{ref:Mishra} raises again the question of the role of the PG
in the high-temperature superconducting cuprates \cite{ref:Norman}. The
elimination of the BCS instability is consistent with the view that the PG
competes with superconductivity. However, as noted in [1], the onset of
superconductivity with a $T_c \sim 90$ K suggests an alternative scenario in
which the PG reflects the formation of short range pairing correlations. Here,
we report results obtained from a dynamic cluster quantum Monte Carlo
approximation (DCA) for a 2D Hubbard model and conclude that (1) the PG, like
the superconductivity, arises due to short-range antiferromagnetic
correlations and (2) contrary to the usual case in which the pairing
instability arises from the Cooper instability, here, the strength of the
spin-fluctuations increases as the temperature decreases leading to the
pairing instability. }
\\

The superconducting transition temperature can be determined from the Bethe-Salpeter
gap equation
\begin{align}\label{eq:1}
-\frac{T}{N}\sum_{n'k'}&\Gamma^{pp}_{\rm irr}(k,\omega_n,k',\omega_{n'})G
(k',\omega_{n'})
 G(-k',-\omega_{n'})\\\nonumber
 &\times\phi_\alpha(k',\omega_{n'})=\lambda_\alpha\phi_\alpha(k,\omega_n).
\end{align}
Here $G(k,\omega_n)$ is the dressed single particle Green's function,
$\Gamma^{pp}_{\rm irr}$ the irreducible particle-particle pairing vertex and $k$
and
$\omega_n=(2n+1)\pi T$ are the usual momentum and Matsubara frequencies,
respectively. The temperature at which the leading eigenvalue of
Eq.~(\ref{eq:1}) goes to 1 gives $T_c$ and the corresponding eigenfunction
$\phi_\alpha(k,\omega_n)$ determines the symmetry of the gap. In spin
fluctuation theories the pairing vertex is approximated by an effective
interaction
\begin{equation}
V_{\rm eff}(q,\omega_m)=\frac{3}{2}\bar U^2\chi(q,\omega_m)
\label{eq:2}
\end{equation}
with $\chi(q,\omega_m)$ the spin susceptibility and $\bar{U}$ a coupling
strength. Various groups have used experimental
data to model $\chi(q,\omega_m)$, $G(k,\omega_n)$ and $\bar U$ in order to
determine whether a spin-fluctuation pairing interaction is consistent with the
observed $T_c$ values.

Dahm et al. \cite{ref:Dahm} used inelastic neutron scattering (INS)
measurements for YBa$_2$Cu$_3$O$_{6.6}$ to model the spin susceptibility
$\chi(q,\omega_m)$ and a one-loop self-energy approximation to determine $G$.
$\bar U$ was an adjustable parameter estimated from INS and ARPES data. Using
the resulting $G$ and $V_{\rm eff}$ in Eq.~(\ref{eq:1}), they concluded that a
spin-fluctuation interaction had sufficient strength to account for the
observed $T_c$. Nishiyama \etal\ \cite{ref:Nishiyama} used inelastic neutron
scattering results for $\chi(q,\omega)$ and solved the Eliashberg equations
for the heavy fermion compounds CeCuSi$_2$ and CeIrIn$_3$. For reasonable
values of $\bar U$, they found $T_c$ values which were again consistent with
the notion that antiferromagnetic spin fluctuations were responsible for
pairing in these materials. In a recent paper Mishra, \etal  \cite{ref:Mishra}
used angular resolved photoemission spectroscopy (ARPES) data for a slightly
underdoped BSCCO ($T_c=90$K) sample to examine the effect of the pseudogap
(PG) on the superconducting transition temperature and to determine whether a
spin-fluctuation pairing mechanism could account for the observed $T_c$. They
found that the usual BCS logarithmic divergence associated with the
propagators in Eq.~ (\ref{eq:1}) was destroyed by the pseudogap and the
leading eigenvalue $\lambda_d(T)$ remained small, and was essentially
independent of temperature. This raises old questions regarding the interplay
between the PG and superconductivity \cite{ref:Norman} which continue to be of
interest \cite{ref:Sordi,ref:Chen,ref:Gunnarsson,ref:Gull13}. Here, using the
dynamic cluster approximation (DCA), we explore spin-fluctuation pairing in a
Hubbard model which exhibits a PG.

The two-dimensional Hubbard model we will consider has a near neighbor hopping
$t$, a next near neighbor hopping $t'/t=-0.15$, an onsite Coulomb interaction
$U/t=7$ and a filling $\langle n\rangle=0.92$. We will work in energy units
where $t=1$. The DCA calculations \cite{ref:RevModPhys77-1027-2005} were
carried out on a $4\times 4$ cluster and employed both continuous-time,
auxiliary-field (CT-AUX) quantum Monte Carlo (QMC) \cite {ref:Gull} and
Hirsch-Fye (HF) QMC \cite{ref:HF} methods to solve the effective cluster
problem \cite{ref:QMC}. In the DCA approximation, where $\Gamma^{pp}_
{\rm
irr}$ depends only on a finite set of cluster momenta $K$, the $k$-sum 
in Eq.~(\ref{eq:1}) gives \cite{ref:Maier06}
\begin{align} \label{eq:DCABSE}
-\frac{T}{N_c}\sum_{n',K'} &\Gamma^{pp}_{\rm irr}(K,\omega_n,K',\omega_{n'})
\bar{\chi}_0^{pp}(K',\omega_n') \phi_\alpha(K',\omega_{n'})\\\nonumber
&  = \lambda_\alpha \phi_\alpha(K,\omega_n)\,.
\end{align}
Here $N_c=16$ is the cluster size and the pairing kernel $G(k,\omega_n)G
(-k,-\omega_n)$ has been coarse-grained (averaged) over the momenta $k'$ of
the DCA patches 
\begin{align}\label{eq:chi0PP}
\bar {\chi}_0^ {pp} (K,\omega_n)=\frac{N_c}{N}\sum_
{k'}G(K+k',\omega_n)G(-K-k',-\omega_{n'})\,.
\end{align} 

\begin{figure}[htbp]
\includegraphics[width=8cm]{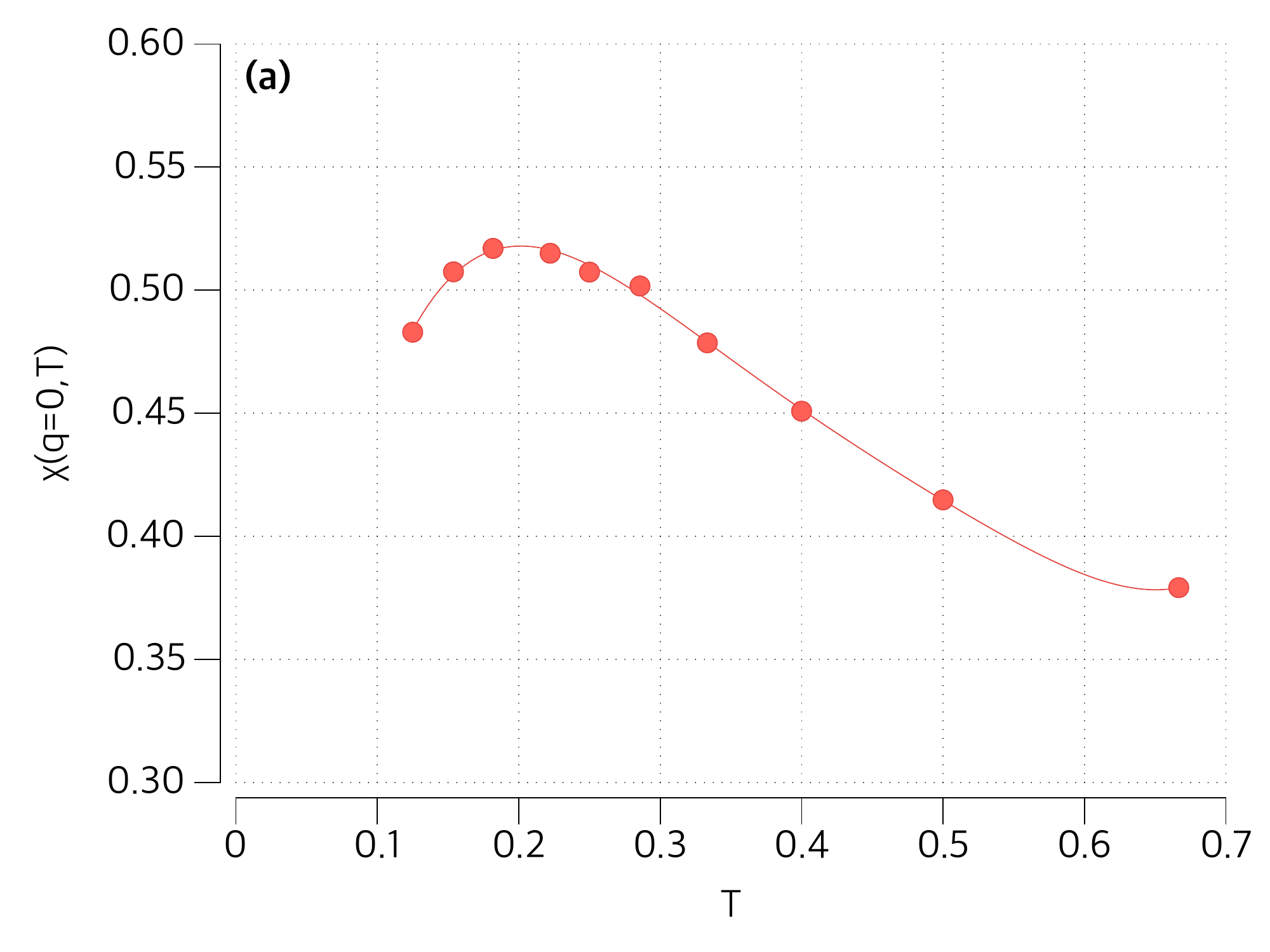}
\includegraphics[width=8cm]{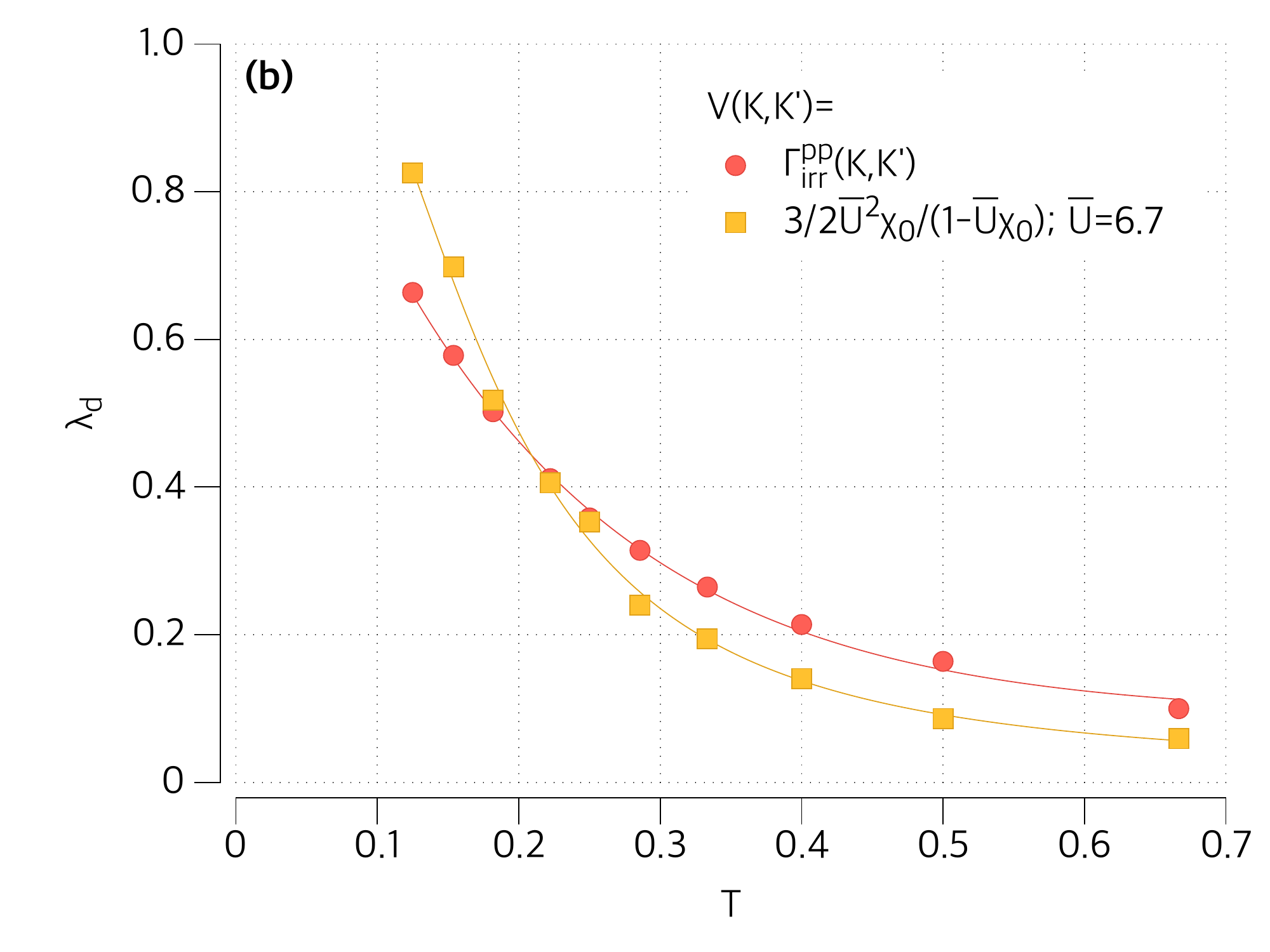}
\caption{{\bf Pairing in the presence of a pseudogap.} (a) The uniform static
spin susceptibility $\chi(q=0,T)$ versus temperature for $\langle
n\rangle=0.92$ $t'=-0.15$ and $U=7$ peaks at a temperature $T^*=0.22$ and
decreases below this as the pseudogap opens. (b) The leading eigenvalue
$\lambda_d(T)$ of the particle- particle Bethe- Salpeter equation versus
temperature (circles) from a DCA calculation of $\Gamma^{pp}_{\rm irr}$. The
$d$-wave eigenvalue for the spin- fluctuation interaction with
$\chi(q,\omega_m)$ the RPA spin susceptibility from Eq.~\ref {eq:chiRPA} and
$\bar{U}=6.7$ is shown as the solid squares.
\label{fig:1}}
\end{figure}

For the parameters we have chosen, the uniform static
susceptibility $\chi (q=0,T)$ versus temperature, shown in Fig.~\ref{fig:1}a,
exhibits a peak at $T^*=0.22$ below which it decreases as $T$ is reduced
\cite{ref:Huscroft}. This behavior, seen in measurements of the magnetic
susceptibility \cite{ref:Johnston} and Knight shifts \cite{ref:Alloul} of
underdoped (hole) cuprates, reflects the opening of a pseudogap. ARPES
experiments \cite{ref:Marshall,ref:Norman98} find that this gap is
anisotropic, opening in the antinodal regions of the Fermi surface. This
behavior has also been seen in DCA calculations of the single-particle
spectral weight \cite{ref:Huscroft,ref:Gull10}. In Fig.~\ref {fig:1}b, the
temperature dependence of the leading eigenvalue of the Bethe-Salpeter
equation (\ref{eq:DCABSE}) is shown as the circles. Its eigenfunction has
$d$-wave symmetry and $\lambda_d(T)$ approaches 1 at low temperatures. Thus
this model system has a pseudogap that opens below $T^*$ and a $d$-wave
eigenvalue that increases towards 1 as $T$ decreases.

\begin{figure}[htbp]
\includegraphics[width=8cm]{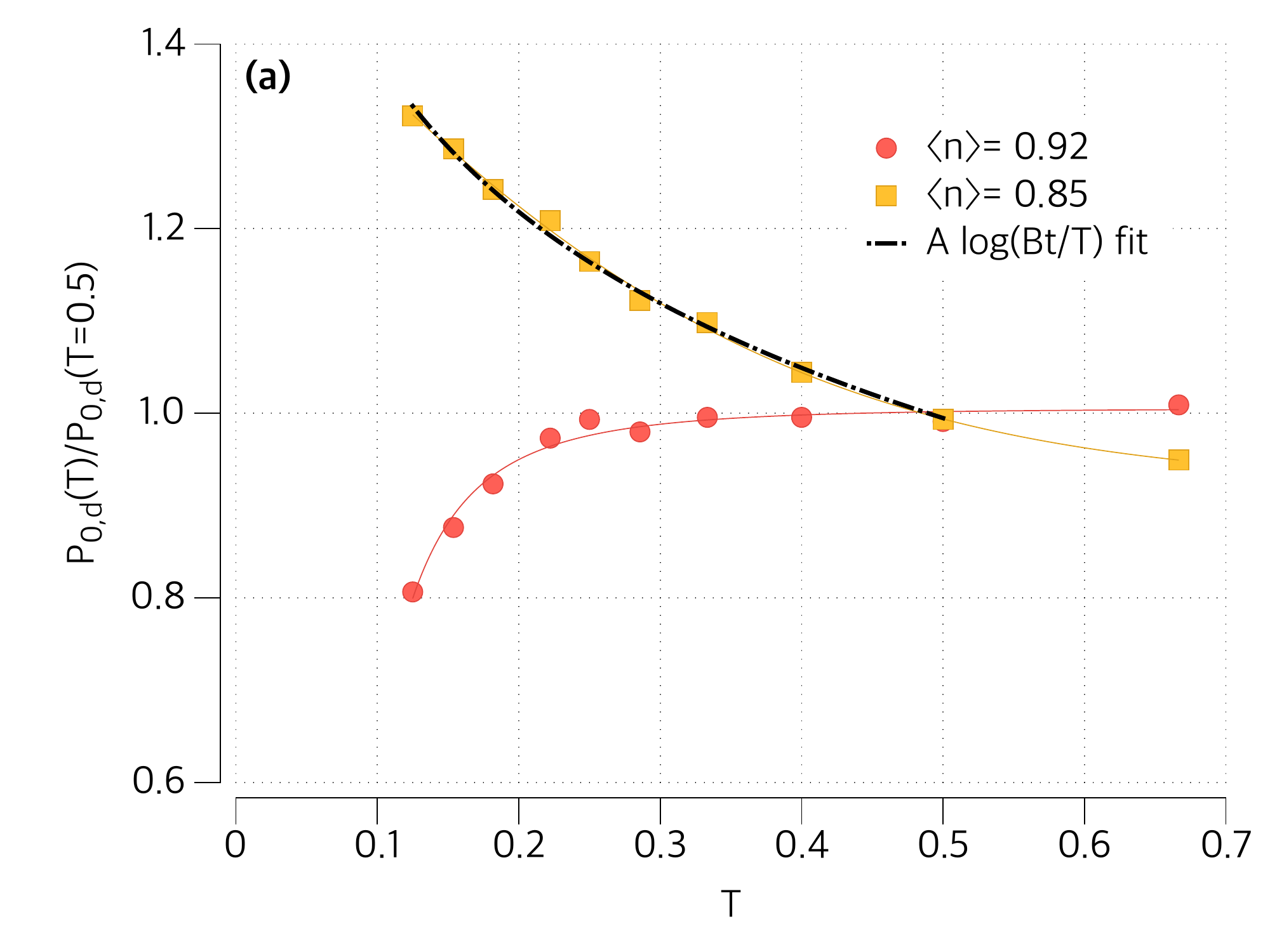}
\includegraphics[width=8cm]{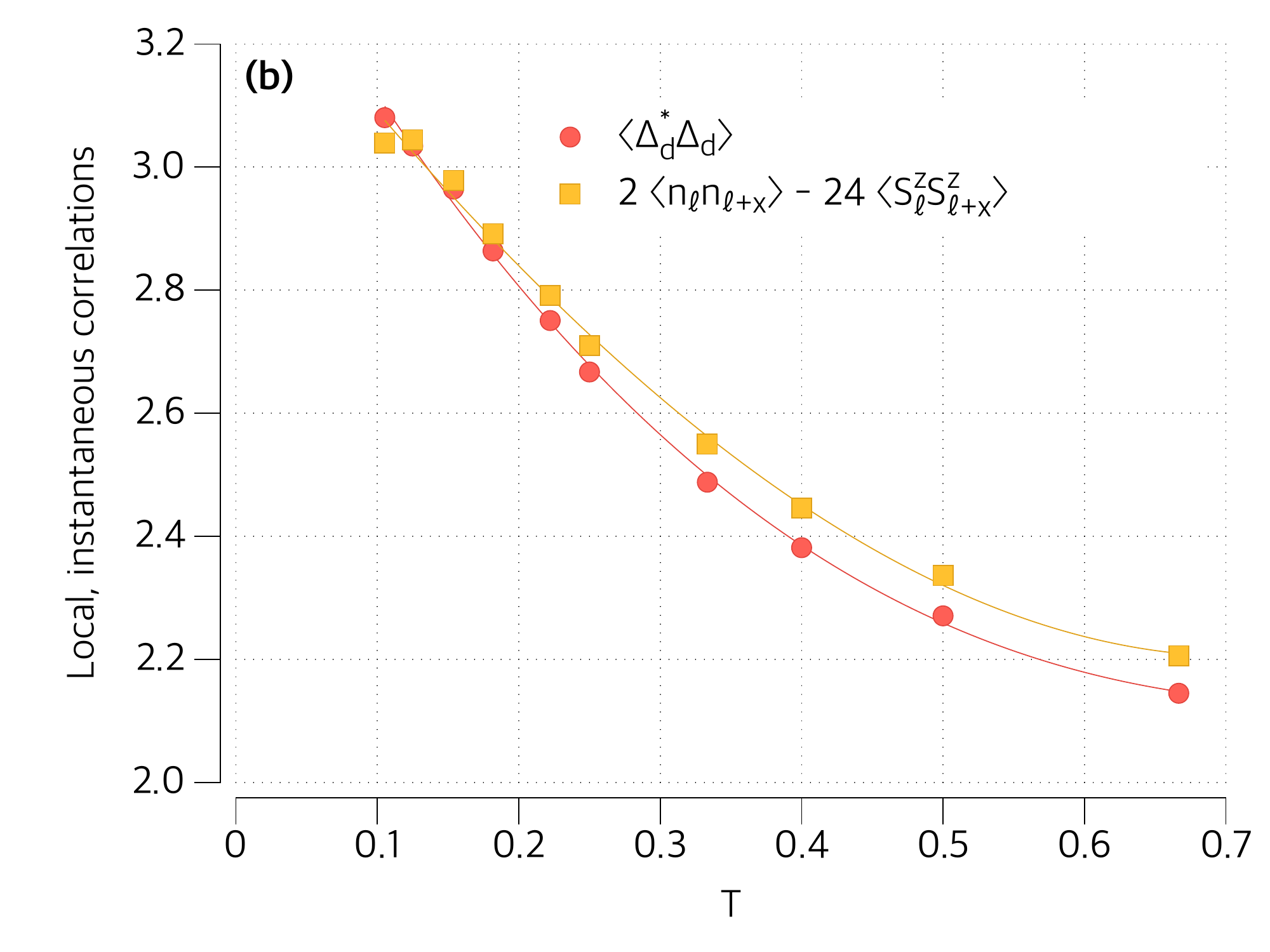}
\caption{{\bf Destructure of the BCS logarithmic instability and nature of
local pairing correlations.} (a) The logarithmic BCS increase of the $d$-wave
projection of the pairing kernel $P_{0d}(T)$ for $\langle n\rangle=0.92$ is
suppressed by the opening of the pseudogap (circles). Here $P_{0d}(T)$ has
been normalized to 1 at a temperature $T=0.5t$ above $T^*$. At temperatures
below $T^*$, where the pseudogap has opened, the BCS logarithmic divergence is
suppressed. The solid squares show $P_{0d}(T)$ for a filling $\langle
n\rangle=0.85$ where there is no pseudogap and one sees the usual logarithmic
increase as the temperature decreases. (b) The temperature dependence of the
local $d$-wave pairfield correlation function
$\langle\Delta^+_d\Delta_d\rangle$ (circles). The observed increase in
$\langle\Delta^+_d\Delta_d\rangle$ as $T$ decreases below $T^*$ reflects the
development of near neighbor AF correlations (squares).
\label{fig:2}}
\end{figure}

In addition to suppressing the $q=0$ spin susceptibility, we find that the
opening of the pseudogap destroys the low temperature BCS logarithmic
divergence of the $d$-wave projection of the pairing kernel
\begin{equation}
  P_{0d}(T)=-\frac{T}{N_c}\sum_{K,\omega_n}\phi_d(K,\omega_n)\bar{\chi}_0^
  {pp}(K,\omega_n)\phi_d (K,\omega_n)\label{eq:3}
\end{equation} Here, $\bar{\chi}_0^
  {pp}(K,\omega_n)$ is defined in Eq.~(\ref{eq:chi0PP}) and
  $\phi_d(k,\omega_n)$ is the $d$-wave eigenfunction, which is approximated as
\begin{equation}
  \phi_d(k,\omega_n)\sim\begin{cases}(\cos k_x-\cos k_y)&|\omega_n|<J\\
	0&\mbox{otherwise}\end{cases}\label{eq:4}
\end{equation} 
with $J\sim4t^2/U$. A plot of $P_{0d}(T)$ versus $T$ is shown
in Fig.~\ref{fig:2}a
 and one can see that below $T^*$, $P_{0d}(T)$ is suppressed as
the pseudogap opens \cite{ref:Kyung,ref:Yang}. Here we have normalized $P_{0d}(T)$ to
its value at a temperature $T=0.5t$ above $T^*$. For comparison, the solid
squares in Fig.~\ref{fig:2}a show $P_{0d}(T)$ for $\langle n\rangle=0.85$
which does not have a pseudogap and one sees the usual BCS logarithmic
behavior (dashed curve).

\begin{figure}[htbp]
\includegraphics[width=8cm]{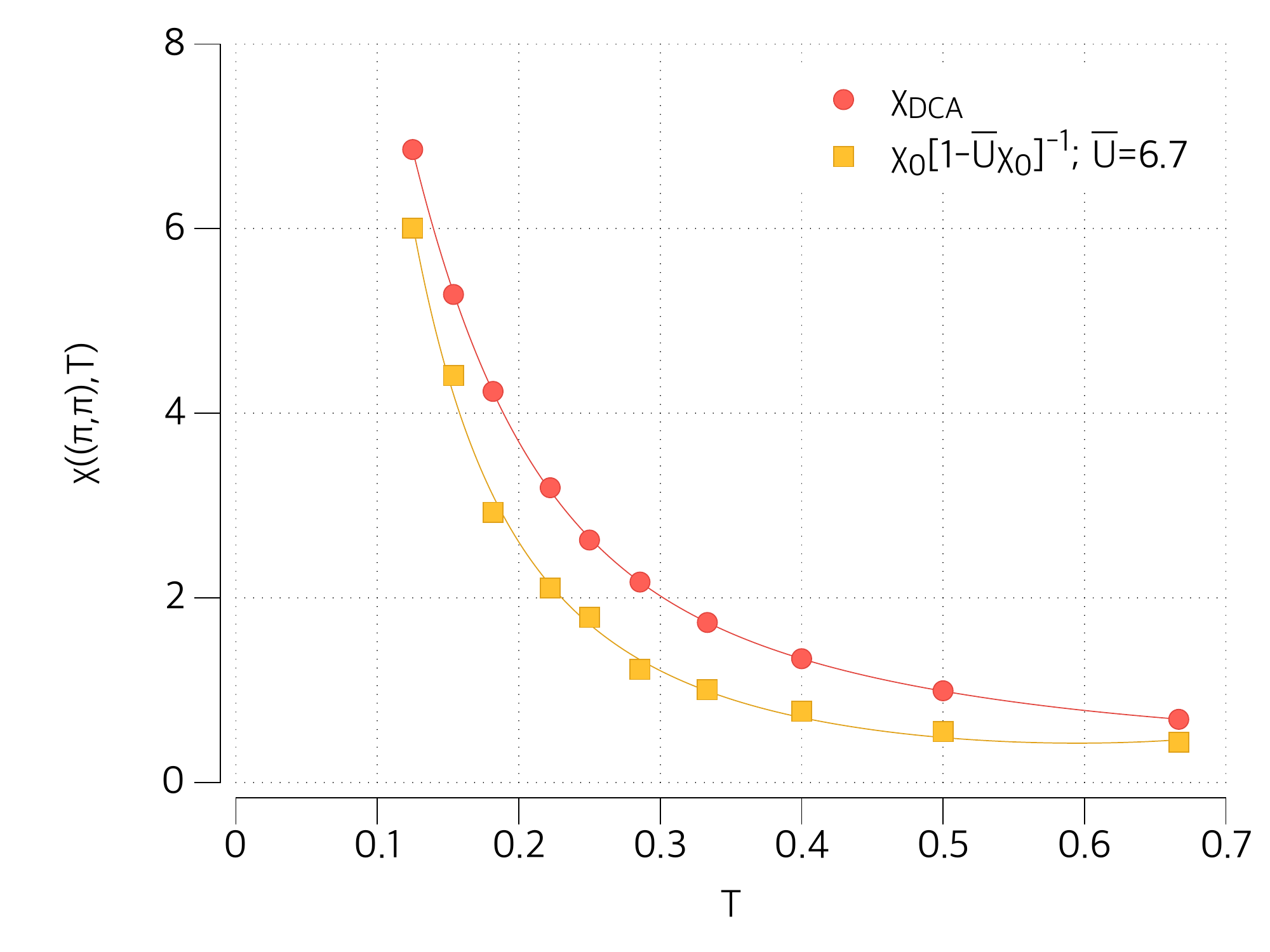}
\caption{{\bf DCA spin susceptibility and RPA fit.} The AF spin susceptibility
$\chi_{\rm DCA}(Q= (\pi,\pi),\omega_m=0)$ from the DCA calculation (circles)
and the RPA fit, Eq.~(\ref{eq:chiRPA}), with $\bar{U}=6.7$ (squares). The AF
response continues to increase as $T$ decreases below $T^*$ leading to an
increase of the spin- fluctuation interaction so that even though the BCS
logarithmic increase of $P_0(T)$ is suppressed, the $d$-wave eigenvalue
$\lambda_d(T)$ increases as seen in Fig.~\protect{\ref{fig:2}}.
\label{fig:3}}
\end{figure}

The absence of the BCS divergence in $P_{0d}(T)$ when there is a pseudogap is
consistent with the finding of Mishra {\it et al.} \cite{ref:P0dT}. However,
as noted, they found that with this suppression, the spin-fluctuation pairing
interaction failed to give a superconducting transition. Based on this, they
suggested that the pseudogap reflects the presence of short-range pairfield
correlations which grow below $T^*$ and become coherent at $T_c$. This
behavior could be likened to the magnetic response of the large $U$ half-
filled Hubbard model. In this case, the formation of local moments when the
temperature drops below $\sim U/2$ is seen in an increase in the expectation
value of the square of the local moment $\langle
S^2_z\rangle=\langle\left(\frac{1}{2}
\left(n_\uparrow-n_\downarrow\right)\right)^2\rangle$. In a similar way one
can look for the onset of local pair formation as $T$ decreases below the
pseudogap temperature $T^*$. Here with
$\Delta^\dagger_{\ell+x,\ell}=c^\dagger_ {\ell+x\uparrow} c^\dagger_{\ell\downarrow}
-c^\dagger_{\ell+x\downarrow}c^\dagger_{\ell\uparrow}$ and
$\Delta^\dagger_d=(\Delta^\dagger_{\ell+x,\ell}-
\Delta^\dagger_{\ell+y,\ell}+\Delta^\dagger_{\ell-x,\ell}-\Delta^\dagger_
{\ell-y,\ell})$, we have calculated
$\langle\Delta^\dagger_d\Delta^{\phantom\dagger}_d\rangle$ versus temperature.
As shown in Fig.~\ref{fig:2}b, this correlation function does increase as the
temperature decreases. However, the four near neighbor pairfield correlations
\begin{equation}
  \langle\Delta^\dagger_{\ell+x,\ell}\Delta^{\phantom\dagger}_
  {\ell+x,\ell}\rangle=\frac {1} {2}\langle n_\ell n_{\ell+x}\rangle-6\langle
  S^z_\ell S^z_{\ell+x}\rangle,\label{eq:5}
\end{equation} contribute the dominant contribution to this increase as shown
in Fig.~\ref{fig:2}b. These results suggest that the pseudogap is more closely
related to the formation of short range antiferromagnetic correlations than to
local pair correlations in agreement with earlier ideas of Johnston
\cite{ref:Johnston} and more recent theoretical results
\cite{ref:Sakai,ref:Sordi,ref:Chen,ref:Gunnarsson,ref:Gull13}.
This identification of the PG with the development of short-range AF spin
correlations is also consistent with the increase of the spin-susceptibility
$\chi(Q=(\pi,\pi),\omega_m=0)$ as shown in Fig.~\ref{fig:3} and as seen
experimentally \cite{ref:Ouazi}.

Returning to the question of whether the spin-fluctuation interaction, Eq.
(\ref{eq:2}), can lead to superconductivity when the logarithmic singularity of
the BCS kernel is suppressed, we use DCA results for $G(k,\omega_n)$ to
construct $V_{\rm eff}(q,\omega_m)$. Here, following Mishra {\it et al.}, an RPA
form for $\chi$ is used
\begin{align} \label{eq:chiRPA}
\chi_{\rm RPA}(Q,\omega_m) = \frac{\chi_{0}(Q,\omega_m)}{1-\bar{U}\chi_{0}(Q,\omega_m)}
\end{align}
with 
\begin{align}
\chi_{0}(Q,\omega_m) = \frac{T}{N_c}\sum_K \bar{G}
(K+Q,\omega_n+\omega_m)\bar {G} (K,\omega_n)\,,
\end{align}
where $\bar{G}(K,\omega_n)=N_c/N\sum_{k'}G(K+k',\omega_n)$ is the DCA
coarse-grained Green's function. The coupling $\bar{U}$ is estimated from the
approximate fit of $\chi_{\rm RPA}$ to $\chi_{\rm DCA}$ shown in Fig.~\ref
{fig:3}.

Then, replacing $\Gamma^{pp}_{\rm irr}$ by $V_ {\rm eff}$ and using DCA
Green's functions, we solve the Bethe-Salpeter equation (\ref{eq:DCABSE}).
Results for $\lambda_d(T)$ are shown (solid squares) in Fig.~\ref {fig:2}. We
conclude that the increase in the strength of the pairing interaction $V_{\rm
eff}$ leads to an increasing $\lambda_d(T)$ similar to that which is found
using $\Gamma^{pp}_{\rm irr}$ determined from the DCA calculation. Thus, in
spite of the absence of the BCS logarithmic increase in $P_{0d}(T)$, we find
that  the increase in the strength of the spin-fluctuations leads to an
increase in $\lambda_d(T)$ as the temperature is lowered. This differs from
the results of reference \cite{ref:Mishra} and we speculate that this
difference  arises from a failure of their parametrization of $G(k,\omega_n)$
by ARPES data taken at 140 K as the temperature is lowered.

To summarize, we have used DCA calculations for an under (hole) doped 2D Hubbard
model, which exhibits a pseudogap, to see whether a spin-fluctuation interaction
provides a reasonable approximation of the irreducible pairing interaction. In
this calculation, the dynamic mean-field cluster is such that charge density and
striping instabilities are suppressed, leaving antiferromagnetic and $d$-wave
pairing as the dominant correlations. We find that while the pseudogap
eliminates the
usual BCS logarithmic divergence of the pairing kernel, a pairing
instability arises from an increase in the strength of the spin-fluctuation
interaction as the temperature decreases. 
% In addition, we find that the
% pseudogap is associated with the presence of short-range antiferromagnetic
% correlations. 

\section*{Acknowledgments}

The authors want to thank V. Mishra and M.R. Norman for useful discussions and
for sending them a plot of $P_{0d}(T)$ calculated using their ARPES derived
single particle Greens function. The authors also want to thank E.~Gull,
S.~R.~Kivelson, A.-M.~Tremblay, and L.~Taillefer for useful comments. DJS and
TAM acknowledge the support of the Center for Nanophase Materials Science at
ORNL, which is sponsored by the Division of Scientific User Facilities, U.S.
DOE. An award of computer time was provided by the Innovative and Novel
Computational Impact on Theory and Experiment (INCITE) program. This research
used resources of the Oak Ridge Leadership Computing Facility, which is a DOE
Office of Science User Facility supported under Contract DE- AC05-00OR22725.

%This is a sample list of references:


\begin{thebibliography}{99}
\bibitem{ref:Mishra} Vivek Mishra, U. Chatterjee, J. C. Campuzano, and M. R. Norman, {\it Nat. Phys.} {\bf 10}, 357 (2014).

\bibitem{ref:Norman} M. R. Norman, D. Pines, C. Kallin, {\it Adv. Phys.} {\bf 54}, 715 (2005).

\bibitem{ref:Dahm} T. Dahm \etal, {\it Nat. Phys.} {\bf 5}, 217 (2009).

\bibitem{ref:Nishiyama} S. Nishiyama, K. Miyake and C.M. Varma, {\it Phys. Rev. B} {\bf 88}, 014510 (2013).

\bibitem{ref:Sordi} G. Sordi, P. S\'emon, K. Haule, A.-M. S. Tremblay, {\it Phys. Rev. Lett.} {\bf 108}, 216401 (2012).

\bibitem{ref:Chen} Kuang-Shing Chen, Zi Yang Meng, Shu-Xiang Yang, Thomas Pruschke, Juana Moreno, Mark Jarrell, {\it Phys. Rev. B} {\bf 88}, 245110 (2013).

\bibitem{ref:Gunnarsson} O. Gunnarsson, T. Sch\"afer, J.P.F. LeBlanc, E. Gull, J. Merino, G. Sangiovanni, G. Rohringer, A. Toschi, arXiv:1411.6947. 

\bibitem{ref:Gull13} E. Gull, O. Parcollet, A. J. Millis, Phys. Rev. Lett.
{\bf 110}, 216405 (2013).


\bibitem{ref:RevModPhys77-1027-2005} Th. Maier, M. Jarrell, Th. Pruschke,
M.H. Hettler, {\it Rev. Mod. Phys.} {\bf 77}, 1027 (2005).

\bibitem{ref:Gull} E. Gull, P. Werner, O. Parcollet, M. Troyer, Europhys. Lett.
{\bf 82}, 57003 (2008).

\bibitem{ref:HF} J. Hirsch, R. Fye, Phys. Rev. Lett. {\bf 56}, 2521 (1986).

\bibitem{ref:QMC} The data in Figs. 1, 2a and 3 was obtained with CT-AUX
QMC and cross-checked with HF QMC. The equal-time data in Fig. 2b was obtained
with HF QMC.

\bibitem{ref:Maier06} T. A. Maier, M. Jarrell, D. Scalapino, Phys. Rev. B.
{\bf 74}, 094513 (2006).

\bibitem{ref:Huscroft} C. Huscroft, M. Jarrell, Th. Maier, S. Moukouri, and A.
N. Tahvildarzadeh, {Phys. Rev. Lett.}, {\bf 86}, 139 (2001). 


\bibitem{ref:Johnston} David C. Johnston, {\it Phys. Rev. Lett.} {\bf 62}, 957 (1989).

\bibitem{ref:Alloul} H. Alloul, T. Ohno, and P. Mendels, Phys. Rev. Lett. {\bf
3}, 1700 (1989).

\bibitem{ref:Marshall} D. S. Marshall, D. S. Dessau, A. G. Loeser, C. H. Park,
A. V. Matsuura, J. N. Eckstein, I. Bozoviz, P. Fournier, A. Kapitulnik,
W.E. Spicer and Z.-X. Shen, {\it Phys. Rev. Lett.} {\bf 76}, 4841 (1996).

\bibitem{ref:Norman98} M.R. Norman, H. Ding, M. Randeria, J.C. Campuzano, T.
Yokoya, T. Takouchi, T. Takahashi, T. Mochiku, K. Kadowaki, P. Guptasarma, and
D. G. Hinks, Nature {\bf 392}, 157 (1998).

\bibitem{ref:Gull10} E. Gull, M. Ferrero, O. Parcollet, A. Georges, A. Millis, Phys. Rev. B. {\bf 82}, 155101 (2010).

\bibitem{ref:Kyung} B.~Kyung, J.-S.~Landry, and A.-M.~S.~Tremblay , PHys. Rev. B {\bf 68}, 174502 (2003).

\bibitem{ref:Yang} S.-X. Yang {\it et al}., Phys. Rev. Lett. {\bf 106}, 047004
(2011).

\bibitem{ref:P0dT} $P_{0d}(T)$ is the $d$-wave
projection of what Mishra \etal\ \protect{\cite{ref:Mishra}} referred to as the pairing kernel. \bibitem{ref:Sakai} T. Sakai and Y. Takahashi, {\it J. Phys. Soc. Jpn.} {\bf 70}, 272 (2001).
\bibitem{ref:Ouazi} S. Ouazi, J. Bobroff, H. Alloul and W.A. MacFarlane, {\it Phys Rev. B} {\bf 70}, 104515 (2004).

%\bibitem{ref:Chan} M. K. Chan, C.J. Dorow, L. Mangin-Thro, Y. Tang, Y. Ge,
%M. J. Veit, X. Zhao, A. D. Christianson, J.T. Park, Y. Sidis, P. Steffens,
%D. L. Abernathy, P. Bourges, M. Greven, arXiv:1402.4472.
%\bibitem{ref:MMJS} T. A. Maier, A. Macridin, M. Jarrell and D. A. Scalapino,
%{\it Phys. Rev. B} {\bf 76}, 144516 (2007).

\end{thebibliography}
\end{document}